# Stability Studies for Photovoltaic Integration using Power Hardware-in-the-Loop Experiments

*Christina N. Papadimitriou[1]\*, Chrysanthos Charalambous[1], Andreas Armenakis[2],*
*Zoran Miletic[3], Werner Tremmel[3], Anja Banjac[3], Thomas I. Strasser[3],*
*Venizelos Efthymiou[1] and George E. Georghiou[1]*

[1] University of Cyprus, FOSS Research Centre for Sustainable Energy, Nicosia, Cyprus
[2] Electricity Authority of Cyprus, Nicosia, Cyprus
[3] AIT Austrian Institute of Technology, Vienna, Austria

\*cpapad11@ucy.ac.cy

## Abstract

The electrical power network is gradually migrating from a centralized generation approach to a decentralized generation with high shares of renewable energy sources (RES). However, power systems with low shares of synchronous generation and consequently low total system inertia, are vulnerable to power imbalances. Such systems can experience frequency stability problems, such as high frequency excursions and higher rates of change of frequency even under small disturbances. This phenomenon is intensified when the grid under investigation has low or no interconnections (islanded) and thus the challenge for stable operation becomes more significant for the operators. This work focuses on how the frequency stability is affected when a photovoltaic (PV) inverter is integrated into a real non- interconnected distribution grid in Cyprus. In order to capture the realistic interactions of this integration, stability experiments in a hardware-in-the-loop (HIL) environment are performed with the aim to provide insightful results for the grid operator.

## 1 Introduction

Power system stability implies the ability of the system to return to normal or stable operation after having been subjected to some form of disturbances. From the classical point of view, power system instability can be seen as loss of synchronism (i.e., some synchronous machines going out of step) when the system is subjected to a particular disturbance [1, 2].

Figure 1 shows the frequency regulation process established by ENTSO-E and presents how the traditional systems are expected to deal with the under-frequency disturbance events within the depicted time-frame and time moment [3]. As shown the system's inertial response is critical since it is the very first and the fastest response available to contain potential instability problems and responsible for achieving/succeeding frequency stability.

As power systems evolve and synchronous units are being replaced by inverter-based technologies (i.e. RES) the overall system inertia lowers [4-6]. Based on this, new challenges related to system stability and robustness are revealed and need to be addressed promptly by system operators [7-9]. This becomes even more accentuated when the grid in question has low interconnections and high rate of RES integration as the inertial response curve (see Figure 1) is greatly supressed downwards.

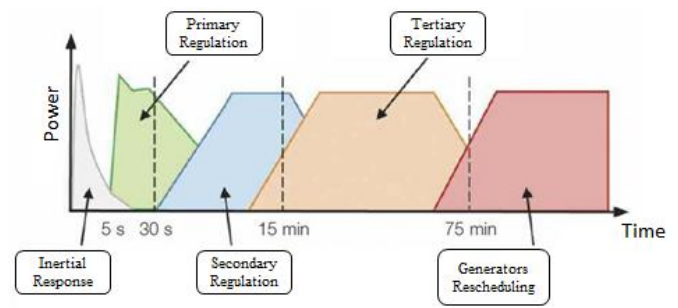

Figure 1. The frequency regulation process steps and timeframe as defined by ENTSO-E [3].

Stability issues in the presence of high shares of RES have already been addressed in the literature. Several authors [10-12] have already studied frequency stability and synthetic inertia provision of PV systems. The authors in [13] investigate the supplementary inertia technologies that could be deployed to mitigate Rate of Change of Frequency (RoCoF) while the authors in [14] present a combination of synthetic inertia (SI) and governor control provided by wind and PV power plants to demonstrate how frequency control can be implemented in bulk power systems.

In real systems, inverter control along with specifications and critical attributes of the power network are taken seriously by the operator before connecting the PV plant. Nevertheless, studying the integration in the field is very difficult due to the following reasons:

- PV is highly dependent on irradiance and temperature
- Full-load tests required by grid codes are not allowed by operators, and hence impossible to be carried out
- High costs associated with field tests in PV
- Lack of flexibility
- Risk of equipment and security of supply.

As a result, offline modelling and simulation tools have been widely introduced to study the power system and its operational challenges such as stability performance upon integration. It has to be stressed, however, that such tools provide insight only on RMS system variables (power, voltages, currents) on certain nodes and power flows between feeders. Furthermore, the grid impacts of solar parks are modelled based on simplified positive sequence EMTP or average "RMS" type models of the studied elements provided by the manufacturers.

In this respect, testing with the help of HIL techniques is of critical importance as it captures the real behaviour of the active assets of the grid without the aforementioned disadvantages of the field testing. Figure 2 shows the general configuration of a PV testing setup along with the power system.

To this extent, in this paper a real-time simulation environment, such as a HIL test platform [15-17], is utilised in order to study frequency stability under a challenging grid environment (i.e., islanded real grid of Cyprus), and under the integration of a large share of PV.

## 2. Methodology

The real-time simulation tests were performed through the HIL framework hosted at the AIT Austrian Institute of Technology premises. These tests reflect a range of different scenarios and realistic test cases of a possible integration of a solar PV park in the power network of Cyprus.

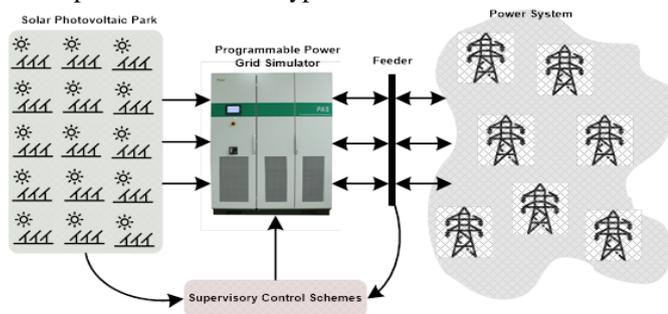

Figure 2: Overview of the HIL configuration synthesizing the solar PV park, electrical feeder interconnection with the power system, and the developed supervisory control schemes ensuring an optimal integration of the PV

### 2.1 Test System Configuration and Equipment

One of the most challenging parts, when integrating RES, is testing of low voltage ride through (LVRT) requirements for inverter compliance. With the help of Controller Hardware in the Loop (CHIL) it is possible to recreate real LVRT according to certain standards. The CHIL simulation is a fitting real-time simulation method that combines numerical simulations, software modelling, and classical hardware testing in laboratories. Control boards represent the hardware device directly connected to the power electronic periphery, which is entirely simulated in real-time simulation conditions. The CHIL approach creates an opportunity to develop the pre-certification capacities in-house and reduce the amount of iterations with an accredited laboratory. The CHIL toolbox enables the iterative development of grid-connected PV inverters [18].

The following lab equipment was employed:

- Typhoon HIL 602+
- AIT Smart Grid Converter (SGC) Controller Card B model
- AIT HIL Controller.

Figure 3 demonstrates the synthesis (set-up) between the Controller Hardware-in-the-Loop (CHIL) based on the Typhoon HIL 602+ hardware infrastructure and Typhoon HIL (v2019.3) software environment, the PV inverter (with control features), modelled on the AIT SGC HIL controller, and the computer unit (user interactive interface). Also, for modelling purposes the following were used:

- Models as Typhoon HIL files
- Time series of the Typhoon HIL scope data
- Scopes from the SGC internal scopes
- Firmware of the SGC, adopted for the solar park.

The solar park PV arrays were modelled according to module published data considering module temperature and irradiation. The PV inverters were modelled with aggregated AIT's SGC PV inverter. The rest of the solar PV park power system components, such as transformers and switch gears, have been modelled according to the plant specifications. The feeder data was modelled according to existing DIgSILENT Power Factory feeder data provided by Electricity Authority of Cyprus (EAC).

In Figure 4, the network model under test is shown. As already mentioned, this is part of the real distribution grid of Cyprus and the highlighted feeder is the one under test for a case where the PV inverter is connected at the end of the feeder. This site was intentionally selected as the PV integration may affect the overall power system stability as well as validating the performance of the developed PV inverter controllers implementing the specific needs of the Grid Rules of Cyprus.

## 2.2 Performance Indicator RoCoF

RoCoF functionalities of protection relays are mandatory for all distributed generators (DG) connected to the grid. This is for securing islanding requirements and thus offer anti-islanding protection to consumers.

According to the Cyprus operator codes, RoCoF relays are mandatory and set to 1.7 Hz/s as the threshold setting to be sustained for 600 msec. If higher values prevail for longer times than 600 ms, the relay operates and isolates the DG since this is the proof for a disconnected system.

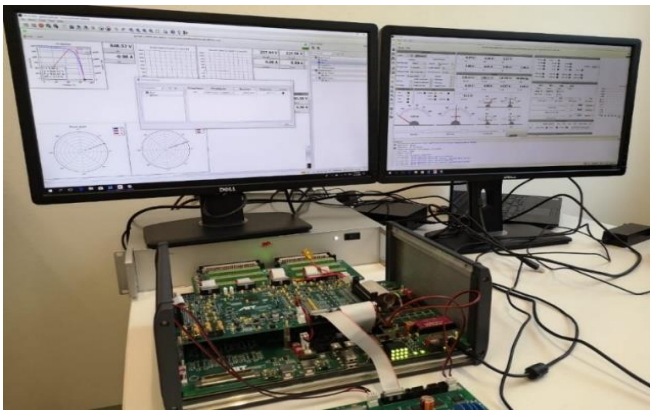

Figure 3: Typhoon HIL (v2019.3) CHIL configuration utilizing a PV power electronic inverter and an intel Core i7-8850H CPU, 2.60GHz, 16GB RAM computer unit

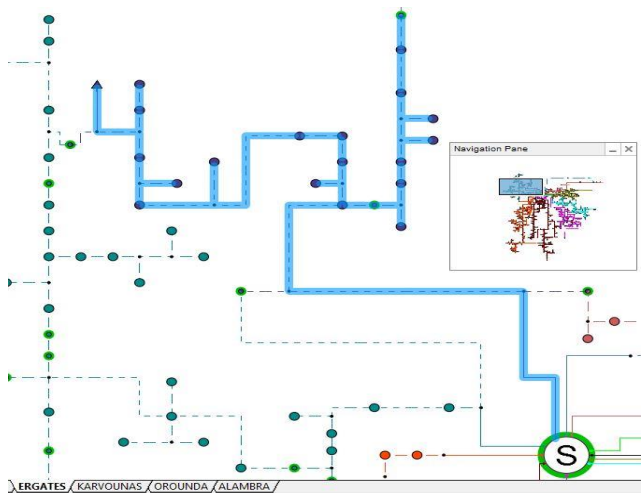

Figure 4: Network model under test

## 3 Results

*3.1 Case Study 1-Voltage Dip under Different PV Generation*

Under this Case study, both transient and steady state stability of the system under different irradiance and thus different PV power generation was tested when a symmetrical voltage dip of 60% occurred at the start of the simulation.

PV plant provides reactive power to support the voltage and at 0.3 sec the voltage is restored. At 0.6 sec, PV plant restores its normal operation, i.e., providing active power and feeding zero reactive power to the grid. (Fig.5-6)

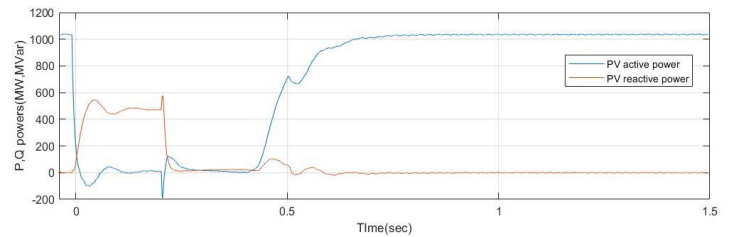

Figure 5: System performance with LVRT functionalities

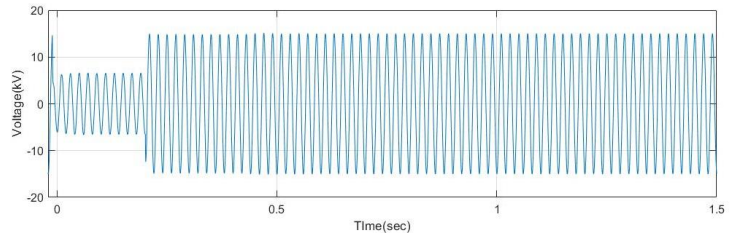

Figure 6: Voltage sag transient and new steady state

As seen in Figures 7-11, the system remains stable throughout the disturbances, being supported by the Low Voltage Ride Through (LVRT) capabilities of the PV inverter as well. All RoCoF settings under different PV generation levels were respected for the grid operator while as expected RoCoF disturbances are more intense when PV generation is higher. In addition, RoCoF at the start, middle and end of the line are captured for all cases.

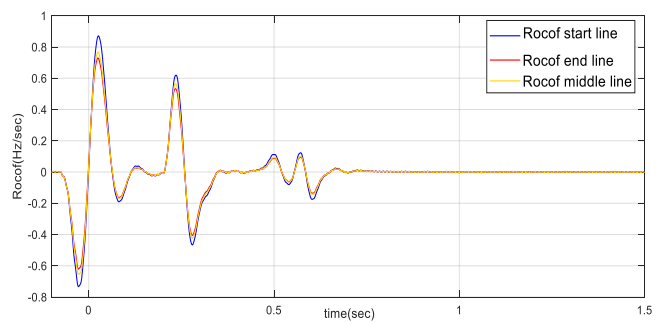

Figure 7. RoCoF under 10% PV generation

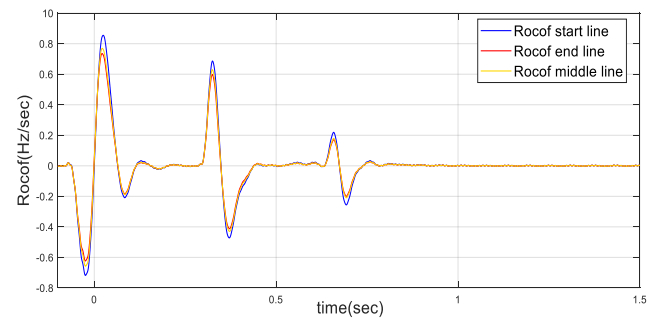

Figure 8. RoCoF under 25% PV generation

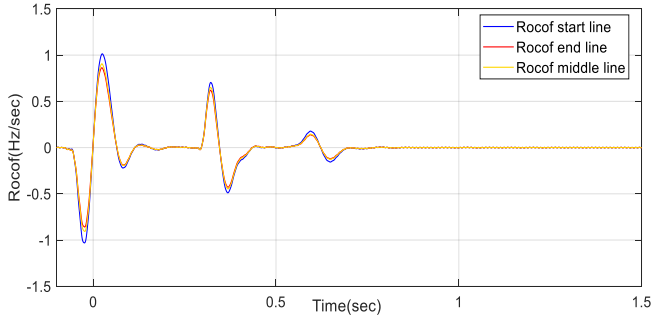

Figure 9. RoCoF under 50% PV generation

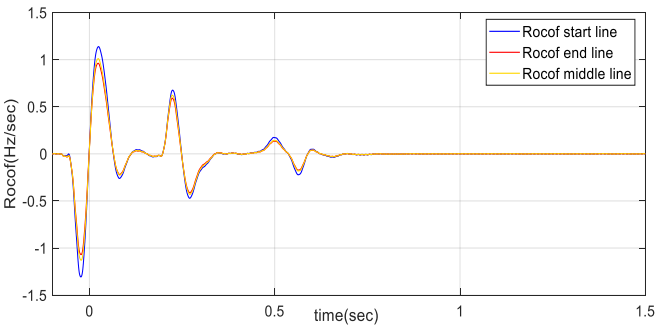

Figure 10. RoCoF under 75% PV generation

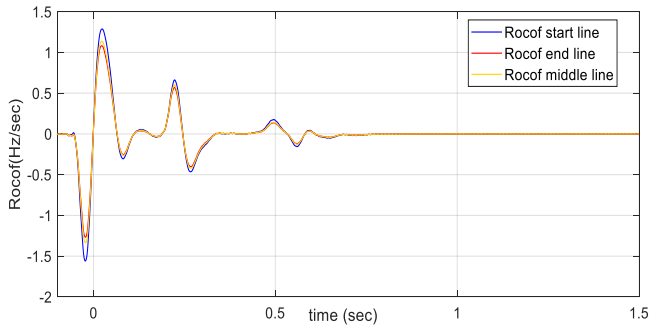

Figure 11. RoCoF under 100% PV generation

### 3.2 Case Study 2-Voltage Dip under Different Loading

The same approach as in the previous case is followed. This time, both transient and steady state stability of the system under different loading was tested when a symmetrical voltage dip of 60% occurred at the start of the simulation time. Again, PV inverter offers LVRT, whereas the PV generation remains the same.

As seen in Figure 12 the system remains stable under all different loadings. It is shown that RoCoF response remains the same for all different loadings. This is expected in the selected loading conditions that were prevailing on the selected system feeding normal passive domestic and commercial loads on the lower end of the low voltage feeders with specific electrical characteristics averaging to approximately 0.9 power factor.

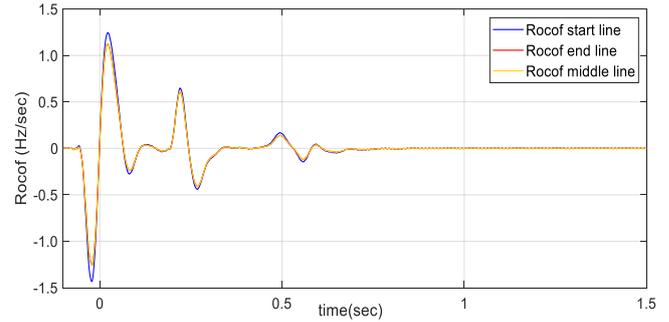

Figure 12. RoCoF under 0, 25,10, 50 and 100% feeder loading

### 3.3 Case Study 3-Voltage Dip under Different Line Lengths

The same approach as in previous cases is followed. This time, both transient and steady state stability of the system under different line lengths was tested when a symmetrical voltage dip of 60% occurred at the start of the simulation.

Figures 13-15 show the RoCoF response under 0.5, 5 and 10-fold longer line length. In all cases, the system stability is secured. As expected, RoCoF is higher under the small disturbances in longer lines due to the bigger impedance. Also, in case of longer lines, RoCoF response differentiate significantly under transients. This also is justified due to the different impedances at the start, end and middle of the long lines.

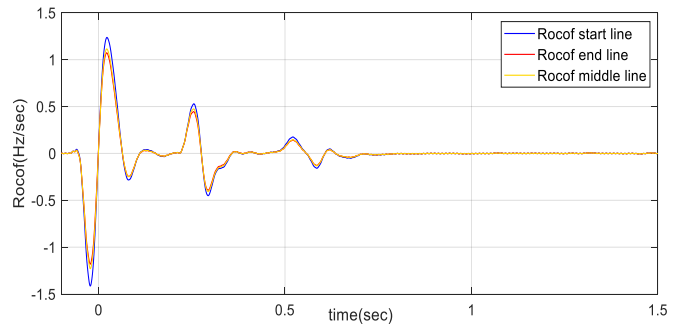

Figure 13. RoCoF with half the original line length

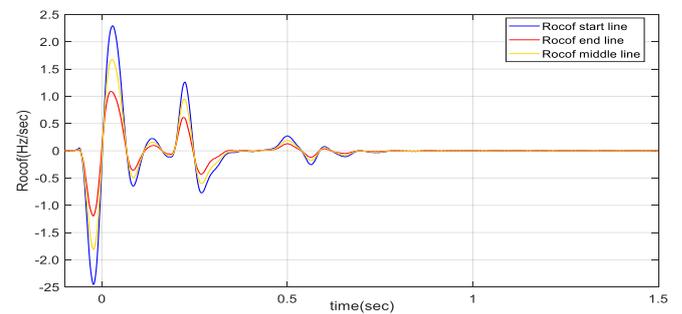

Figure 14. RoCoF with 5-fold longer line length

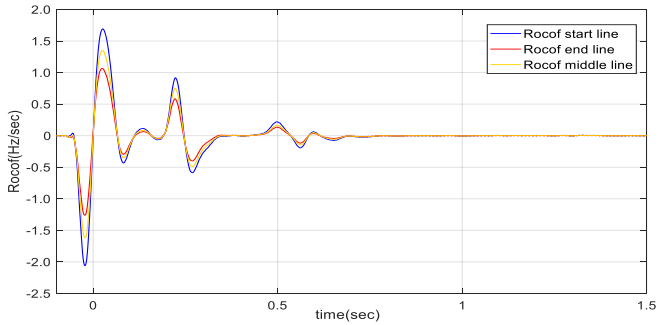

Figure 15. RoCoF with 10-fold longer line length

## 4  Conclusion

High RES penetration is expected to impact the stability of the system as the inherent inertia of the power grid lowers significantly. The challenge of preserving the stability under transients becomes more pressing when the grid is islanded with no interconnections to bulkier systems.

The transient and steady state stability of the Cyprus grid under a serious voltage dip for different system use cases in the presence of a large PV plant integration has been studied in this paper.

PHIL experiments were conducted in order to capture the real interaction between the two systems as the PV inverter employs LVRT capabilities to support the system under the transient event. The main outcomes of this paper are the following:

- The system preserves its stability under different cases of irradiance and hence PV generation. When the PV generation is higher, the RoCoF response is more intense for the same event.
- The system stability is less affected by the loading of the line that the PV is integrated in to.
- The system preserves its stability under different length lines. It is expected that the longer the lines, the bigger RoCoF response is recorded.

Capitalizing on the above findings, it is noted that some open issues for further research can be considered:

- The integration of more than one PV inverter. RoCoF operational needs and response should be tested exhaustively in relation to the provided ancillary services and fault detection and isolation in the cases where system violates the required RoCoF settings provided by the grid rules of Cyprus.
- From available technology point of view, all functionalities apart from LVRT, and classified as ancillary or system services, should be considered in line with the requirements of the prevailing standard.

Finally, in future work, more use cases will be tested in order to investigate in depth the PV integration in existing relatively weak grids for islanded and grid – connected operation.

## 5  Acknowledgements

The authors would like to thank the AIT Austrian Institute of Technology for hosting the Trans-national Access project "CY-PRESS" and providing the lab infrastructure. This work was supported by the European Union's Horizon 2020 grant 654113 (ERIGrid) under the Trans-national Access programme.